# Limit of Field Effect Mobility on Pentacene Single Crystal

V.Y. Butko*, X. Chi, D. V. Lang and A. P. Ramirez
Los Alamos National Laboratory, Los Alamos, New Mexico, USA

We report on fabrication and characterization of field effect transistors (FETs) on single-crystal pentacene.  These FETs exhibit hole conductivity with room temperature effective mobility, $\mu_{eff}$, up to 0.30 cm$^2$/Vs and on/off ratios up to 5×10$^6$. A negative gate voltage of –50V significantly decreases the activation energy ($E_a$) down to 0.143 eV near room temperature. Assuming thermal equilibrium between trapped and free carriers, from $E_a$ = 0.143eV, we find the number of free carriers is only 0.4% of the total number of injected carriers. Along with $\mu_{eff}$ ~ 0.3 cm$^2$/Vs this gives the intrinsic free carrier mobility of ~75 cm$^2$/Vs.

The Field Effect Transistor (FET), based on controlled injection of free carriers into semiconductors, is probably the most prominent constituent of modern microelectronics[1]. Semiconducting organic materials have received increased attention in the last several years because they promise bulk processing of flexible, large-area devices at low cost[2-5]. FETs also provide a powerful method of investigating two-dimensional physical properties of these materials. However, the flexibility of organic materials is also related to the difficulties encountered in reproducing the well-known results obtained in inorganic semiconductors. Unlike inorganic materials, organics possess intrinsic inter- and intra-molecular vibrational modes that have the potential to scatter charge carriers, and reduce mobility, $\mu$, more effectively than usual phonons. Assessing the fundamental limits on free carrier mobility in organic systems is one of the major problems of the physics of organic solids.

Among the most interesting of organic systems are the polyacenes, with pentacene as the most prominent member[2]. Pentacene has the smallest bandgap among the linear polyacenes, and the highest effective mobility in polycrystalline organic thin film FETs, $\mu_{eff}$ = 0.3-1.5 cm$^2$/Vs[5-7]. This value is comparable to that of amorphous Si devices. It is known that $\mu_{eff}$ increases with increasing crystallinity in such thin film transistor (TFT) devices[4,6-8]. This result and tremendous improvement of $\mu$ in inorganic crystals compare to disorder systems motivates the need for results on FETs fabricated from organic single crystals.

In this work we fabricate and study FETs formed on the surface of high-quality pentacene single crystals. These crystals were grown by horizontal physical vapor transport in a stream of ultra high purity argon. The crystal growth apparatus was a modified version of the one reported by Laudise et. al. [9] with two glass tubes of different diameters. The outer one was wrapped with two rope-heaters which define the source zone and crystal growth zone, respectively. A glass tube of smaller diameter serves as the reactor tube. The source temperature was 285 C, and the flow rate was 19 ml /min. The source material was purchased from Aldrich and was twice re-crystallized in argon for purification before used for crystal growth. Typical crystal dimensions are 1-2 mm length, 0.2 – 1 mm width and 0.05 – 0.5 mm thickness. In this work we use two types of source/drain contact configurations: colloidal graphite/colloidal graphite and colloidal graphite/silver paste. Both of these configurations have been painted on the smooth, untreated single crystal surface. The separation between contacts is ~100 μm. The width of the contact pads is ~0.3 mm. Similar to the approach used in[10] on rubrene, the organic parylene was used as gate insulator (400-500 nm thick). Parylene was deposited on the top of the crystal in a home-made reactor. The thickness of the parylene layer is estimated from the capacitance of planar devices fabricated simultaneously with the pentacene devices. In the final step of FET fabrication, a silver paste gate contact electrode was painted on the top of the parylene over the region between the source and drain. No devices have been annealed. The current and voltage were measured with two Keithley 6517A electrometers. The electrical measurements were made in darkness in a Quantum Design cryostat at fixed temperature in a vacuum ~ $10^{-5}$ Torr. For these measurements, a 1-20 Volt step and a 5-60 second delay between each measurement was typical. Leakage gate current at the low voltages was ~$10^{-14}$-$10^{-13}$ A, and at the highest voltages applied, never exceeded $3\times10^{-12}$ A.

Typical transistor characteristics of devices at room temperature are shown in fig.1-2. The drain current ($I_{sd}$) exhibits a linear dependence on source-drain voltage ($V_{sd}$) at fixed gate voltages ($V_g$) (Fig.1) when $V_{sd} < V_g$. For $V_{sd} \geq V_g$ the current saturates at a near-constant value. This is standard behavior of a semiconductor FET[11]. In the right hand side of fig. 2, is shown $I_{sd}$-$V_g$ characteristics at fixed $V_{sd}$. One can see that the on-off ratio reaches $5\times10^6$ at low $V_{sd}$. From the left-hand side of Fig.2, which shows the dependence of $(-I_{sd})^{1/2}$ on $V_g$, we extract an effective mobility $\mu_{eff} \sim 0.30$ cm$^2$/(Vs) at $V_{sd} = -50$ V and threshold voltage $V_t \sim 5$V. Our measurements on a sample with colloidal graphite/silver paste contacts also demonstrate FET action with a few times less mobility.

In Fig. 3 are shown results of a variable-temperature study of a pentacene single crystal FET at different values of $V_g$. We fit these data to the form $R = R_0 \exp(-E_a/T)$, where $E_a$ is an activation energy, and also plot $E_a$ as a function of $V_g$ in fig. 4. We see that for slightly positive $V_g$, in room temperature range $E_{a0} \sim 0.57$ eV, a value approximately 30% of the optical activation energy of $\sim 2$ eV [12]. This is most likely due to an asymmetric distribution of traps which pin the Fermi level toward the valence band. For large negative $V_g$ (-50V), $E_a$ decreases to a significantly lower value, but still a thermally activated transport ($E_{a1} = 0.143$ eV) is observed. This behavior appears to be different from both the TFT results which exclude thermally activated hopping as the fundamental transport mechanism in pentacene thin films[6] and the low temperature results of photoinduced carrier time-of flight (TOF) technique on naphthalene crystals, described by a band model[13]. However, it is very likely that carrier trapping masks the intrinsic transport behavior in our crystals. To reveal the fundamental free carrier crystal mobility, the following analysis can be applied. The observed behavior is well described in terms of a standard semiconductor model assuming that holes injected from the contacts become trapped at and below the Fermi energy level ($E_F$) inside the band gap[14-16]. In this picture $E_a \sim E_F - E_V$ and the dependence of $E_a$ on $V_g$ corresponds to a gradual shift of the hole Fermi energy toward the valence band as more empty shallow[17] traps become filled due to FET hole injection. Therefore the number of injected carriers per unit area

$$C_s V_g \approx \int_{E_{F1}}^{E_{F0}} dE N_t(E) \qquad (1)$$

where $N_t(E)$ is energy distribution function of traps (trapped carriers) per unit area. $E_{F0}$ and $E_{F1}$ are Fermi energy levels without and with $V_g$ on, correspondently. We assume that, at any injection rate, thermal equilibrium is established between free and trapped carriers[14,16]. The number of thermally excited free carriers is given by

$$p_f = \int_{E_{F1}}^{E_C} dE N_t(E) \exp((E_V - E)/k_B T) . \qquad (2)$$

Due to an exponential decrease of the thermal excitation term from the deep trapped carriers, the $E_c$ (conduction band energy) integral limit can be changed to $E_{F0}$ without a significant error at room-temperature. Obtained expression can be overestimated by the following

$$p_f = \exp((E_v - E_{F1})/k_B T) \int_{E_{F1}}^{E_{F0}} dE N_t(E) \approx \exp(-E_{a1}/k_B T) C_s V_g . \qquad (3)$$

From this equation and $E_{a1} = 0.143$ eV at $V_g = -50$ V, we find that at room temperature the number of free carriers is about or less than 0.4% of the total number of injected carriers. This fact along with $\mu_{eff} \sim 0.3$ cm$^2$/Vs gives a free carrier mobility $\mu \sim 75$ cm$^2$/Vs. We note that $\mu_{eff}$ has been studied in the FET configuration in rubrene crystals[10] and $\mu$ by the TOF technique on other crystals[13,18,19]. The FET results on rubrene, $\mu_{eff} \sim 0.1 - 1$ cm$^2$/Vs, are not inconsistent with our present results. The TOF results demonstrate a significant increase of intrinsic mobility above values of $\mu_{eff}$ reported for TFTs. However, a direct confrontation of FET and TOF results for the same material (and same batch) have yet to be made.

As mentioned above, the crystals we studied have a large density of traps. It is conceivable that some of these traps are introduced in the process of FET fabrication, which has not been optimized. Of course, decreasing the number of traps would significantly improve the device characteristics, which remains the ultimate goal of single-crystal device project. On the other hand, this work presents the first fabrication and study of pentancene single crystal FETs and the first experimental indication that room temperature intrinsic pentacene mobility can be as high as 75 cm$^2$/Vs.

We would like to acknowledge useful discussions with J. Orenstein, D. Smith, B. Crone, C. Kloc, V. Podzorov, Gavin Lawes, G. Boebinger and C. Varma. This research was supported by the DOE Office of Basic Energy Sciences.

* On leave from Ioffe Physical Technical Institute, Russian Academy of Science, Russia

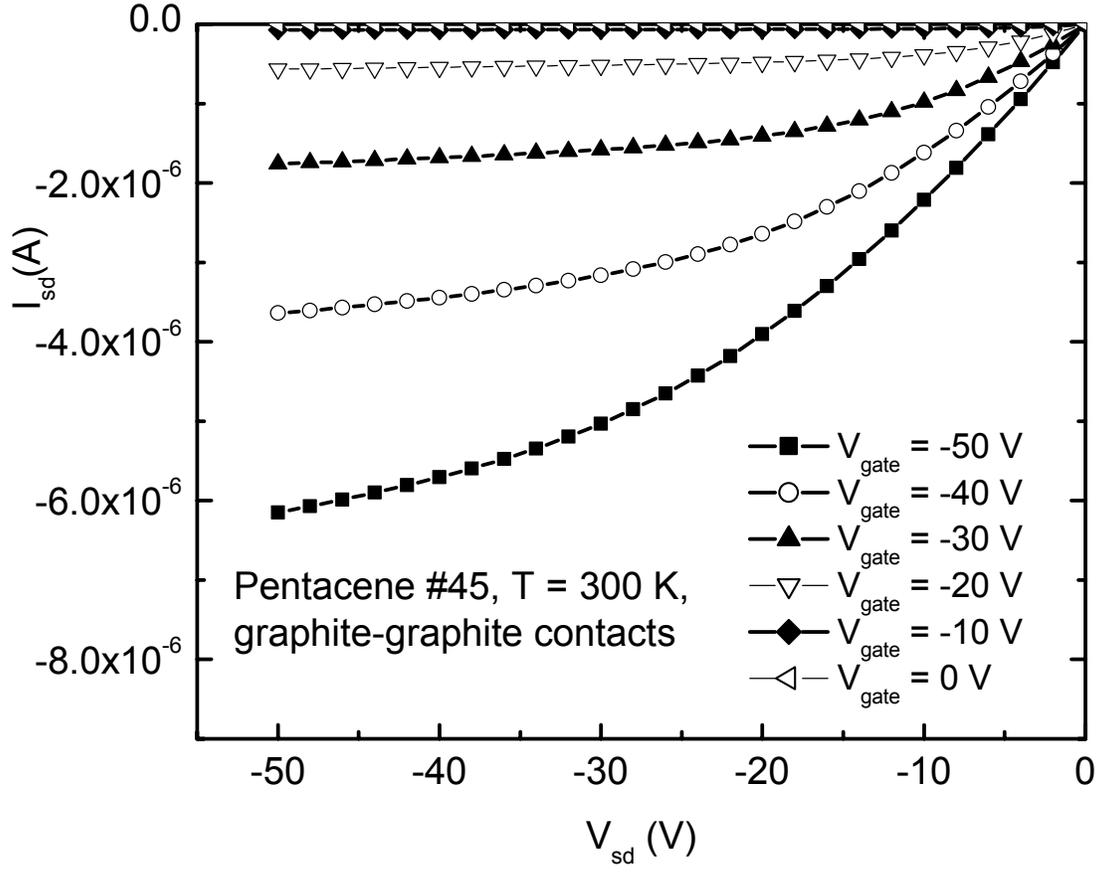

FIG.1.

$I_{sd}$ ($V_{sd}$) FET characteristics at different Vg.

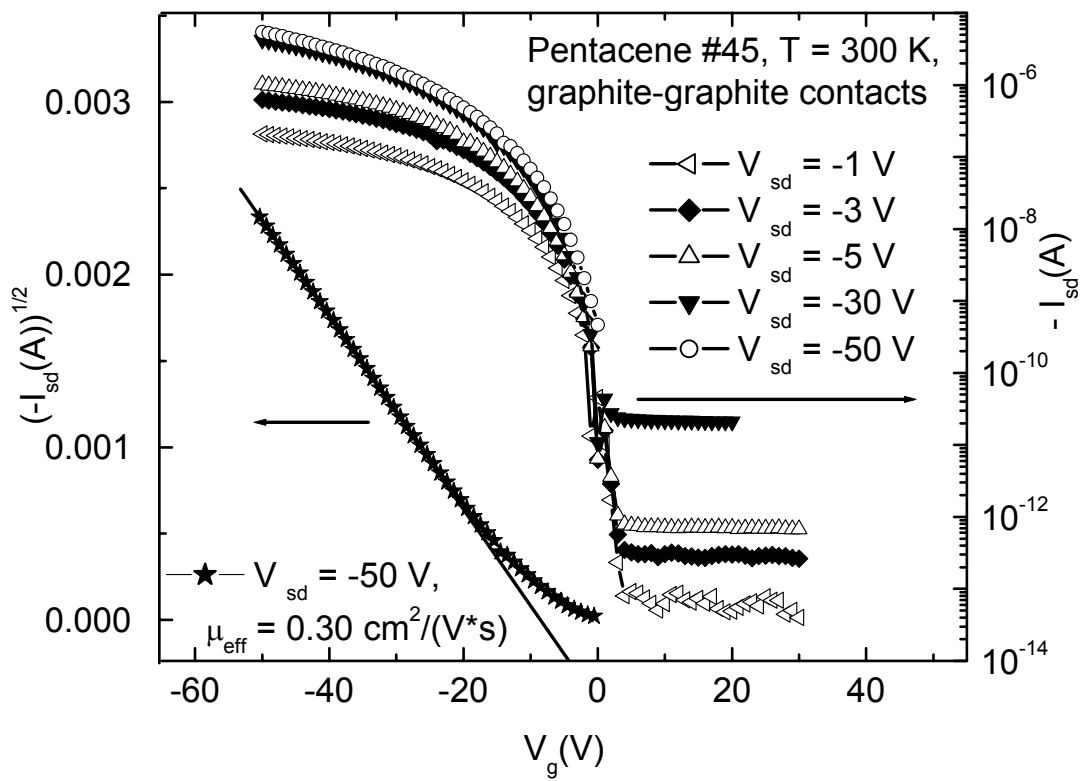

FIG.2.

Right hand side: $-I_{sd}(V_g)$ FET characteristics at different $V_{sd}$.
Left hand side: $(-I_{sd})^{1/2}(V_g)$ FET characteristic at $V_{sd}=-50V$.

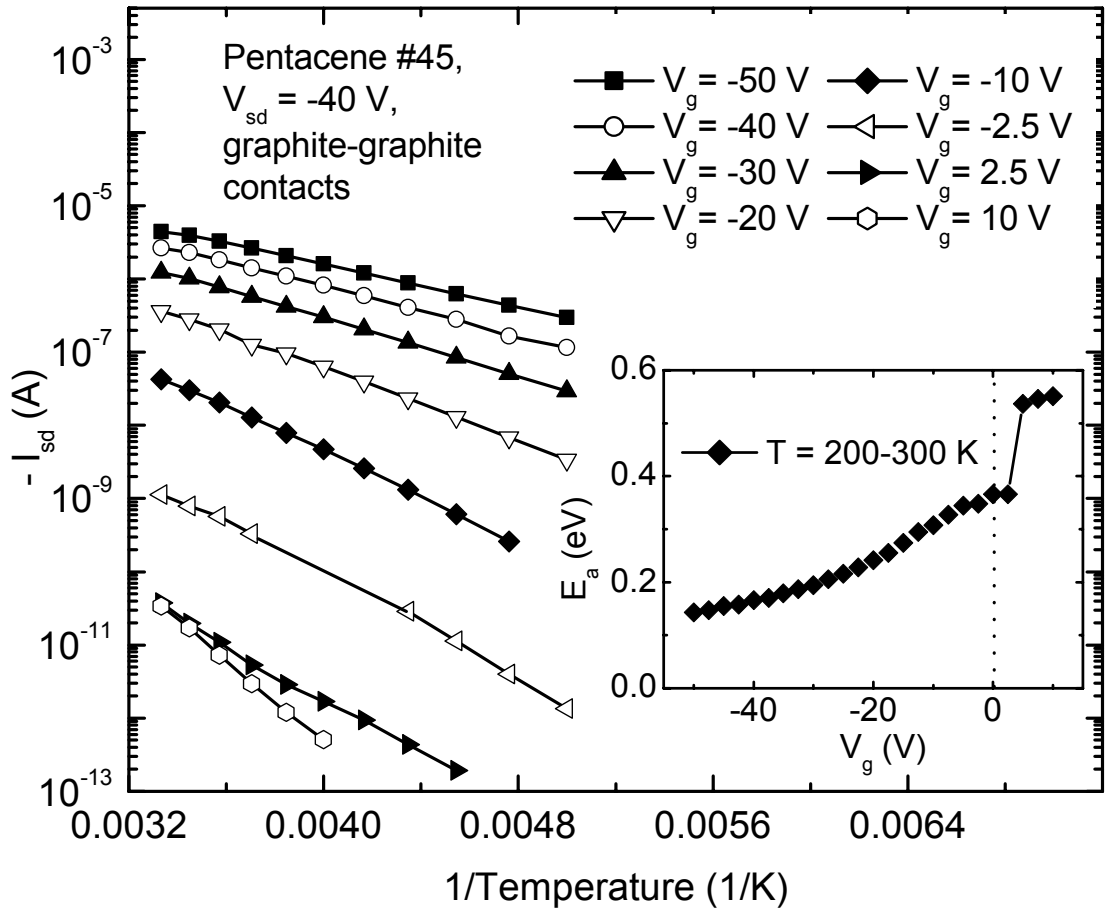

FIG.3.

Main Part: Dependence of $-I_{sd}$ ($V_g$) on 1/Temperature at $V_{sd}$ = -40V
Inset: Dependence of $E_a$ on $V_g$ in 200K-300K temperature range.

9